\documentclass[a4paper,11pt]{article}

\usepackage{amssymb}
\usepackage{amsmath}
\usepackage{amsthm}
\usepackage{mathtools}
\usepackage{epsfig}
\usepackage{dcolumn}
\usepackage{tikz}
\usepackage{tikz-cd}
\usetikzlibrary{shapes.geometric, arrows}
\usepackage{upgreek}
\usepackage{setspace}
\usepackage{enumitem}
\usepackage{array,multirow,bigdelim,arydshln}
\usepackage{appendix}
\usepackage[export]{adjustbox}
\usepackage{xparse}
\usepackage[utf8]{inputenc}
\usepackage{microtype}
\usepackage{bm}
\usepackage{braket}
\usepackage{dsfont}
\usepackage{nccmath}
\usepackage{datetime}

\usepackage{pos}

\def \be {\begin{equation}}
\def \ee {\end{equation}}
\def \nn {\nonumber}
\def \la {\langle}
\def \ra {\rangle}
\def \C {\mathbb{C}}
\def \R {\mathbb{R}}

\def \GL {\mathrm{GL}}

\def \CP {\mathbb{CP}}

\def \P {m}

\def \V {\mathcal{V}}
\def \w {\wedge}
\def \Q {\mathbb{Q}}
\def \dW {\scalebox{.65}{$dW$}}
\def \mdW {\scalebox{.65}{$-dW$}}
\def \pmdW {\scalebox{.65}{$\pm dW$}}
\def \nab {\nabla_{\!\dW}}

\DeclareMathOperator{\Res}{Res}

\newcommand{\overbar}[1]{\mkern 1.5mu\overline{\mkern-1.5mu#1\mkern-1.5mu}\mkern 1.5mu}

\theoremstyle{definition}

\allowdisplaybreaks
\graphicspath{{figures/}}

\title{Status of Intersection Theory and Feynman Integrals}

\author*{Sebastian Mizera}\emailAdd{smizera@ias.edu}
\affiliation{Institute for Advanced Study, Einstein Drive, Princeton, NJ 08540, USA}

\abstract{We give a pedagogical review of the recently-introduced notion of a ``scalar product'' between Feynman integrals and how it helps us understand the analytic structure of the perturbative S-matrix.}

\FullConference{%
  MathemAmplitudes 2019: Intersection Theory \& Feynman Integrals (MA2019)\\
  18-20 December 2019\\
  Padova, Italy}

\tableofcontents

\begin{document}
\maketitle

\section{\label{sec:introduction}Introduction: Gauge Theory on the Kinematic Space}

In the preface to \emph{The Analytic S-Matrix} the authors wrote that ``one of the most remarkable discoveries in elementary particle physics has been that of the existence of the complex plane'' \cite{Eden:1966dnq}. They were not exaggerating. Indeed, questions about physical properties of scattering processes---such as unitarity, causality, crossing symmetry, or dispersion relations---are most cleanly addressed when translated into sharp mathematical statements. While in the context of the original S-matrix program \cite{Eden:1966dnq,Chew:107781} this mostly meant the use of complex analysis, there is no point in kidding ourselves that modern questions in physics would not benefit from the apparatus of more contemporary mathematics. As a matter of fact, already in the 1960's Fotiadi, Froissart, Lascoux, and Pham \cite{FOTIADI1965159,AnalyticStudy1,AnalyticStudy2} realized that algebraic topology plays an important role in the understanding of analytic properties of scattering amplitudes (for reviews and related work, see, e.g., \cite{hwa1966homology,Lascoux:1968bor,Regge:1968rhi,Abreu:2017ptx}). Broadly speaking, it studies how global (topological) aspects of Feynman integrals constrain the sheet structure of Riemann surfaces and their Landau singularities. The point of this article is to summarize a recent set of ideas, collectively referred to as  ``intersection theory'', which can be thought of as a modern twist on these old works.

We do not expect the reader to have any formal training in mathematics. Fortunately, concepts from topology have to certain extent already permeated to physics, especially in the context of gauge theories. Therefore, to keep this article accessible, we will attempt to formulate some of the discussion below in gauge-theoretic terms, hoping that it gives the reader more intuition about the results.

To set the stage, let us briefly review the objects we want to study. We first isolate the trivial contribution to the S-matrix,
$S=\mathds{1} + i T$, and focus on the matrix elements of $T$ between incoming and outgoing states with momenta $p_i$. They are distributions supported on the momentum conservation delta function. For the sake of illustration, let us consider an example of the matrix element for a two-to-two scattering, which can be written as
\be
T_{12 \to 34} = \delta^4(p_1 {+} p_2 {-} p_3 {-} p_4)\, {\cal T}_{12 \to 34}(s,t,p_i^2,m^2,\ldots).
\ee
This process can in principle depend on the two independent Mandelstam invariants $s=(p_1{+}p_2)^2$, $t=(p_2{-}p_3)^2$; masses of external states $p_i^2$; and possibly other mass scales, such as $m^2$, present in our quantum field theory. Let us treat each of these invariants as a complex variable and refer to the space of all their allowed values as the \emph{kinematic space}. A classic question in the S-matrix program asks about the analytic properties of ${\cal T}_{12 \to 34}$ on this space.

We can state the problem of analyticity in the following way, which hopefully makes it suggestive that the aforementioned questions have something to do with topology. Assume that the amplitude ${\cal T}_{12\to 34}$ was evaluated at some specific point $(s_\ast,t_\ast,p_{i\ast}^2,m^2_\ast,\ldots)$ in the kinematic space and then ask what happens to it as we continuously vary the kinematic parameters along a path $\gamma$, for example going around $s{=}4m^2$, in the complex space and return back to the original point:
\be\label{fig:loop}
\includegraphics[scale=1,valign=c]{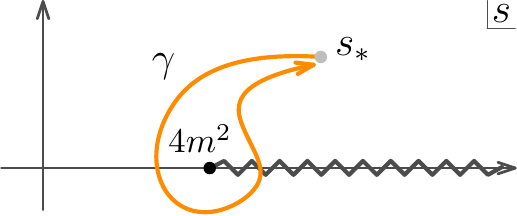}
\ee
Knowing how the result looks like would tell us about any possible branch cuts and discontinuities encountered by $\gamma$. Repeating this exercise for \emph{all possible} loops $\gamma$ (not just those confined to the $s$-plane, as in the example above) amounts to knowing the full analytic structure of the scattering amplitude. Unfortunately, we do not yet have means for addressing this question in full generality.

In order to make progress we will make three simplifications: (a) consider scattering amplitudes in perturbation theory, which allows us to work with Feynman diagrams at a given number of loops, (b) since Feynman integrals often do not converge, we employ dimensional regularization to define them in space-time dimension $4{-}2\varepsilon$ with a small positive parameter $\varepsilon$, and (c) consider families of scalar Feynman diagrams. The last assumption is made so that we can keep our discussion completely theory-agnostic, as an arbitrary Feynman integral can be reduced to a sum of scalar ones. It might be that some of the above assumptions can be lifted for special quantum field theories, but we wish to study the \emph{generic} case in order to encompass S-matrices of realistic theories testable at particle colliders.

While dimensional regularization has been traditionally thought of as a nuisance, more recent work points to it being rather convenient, for example in understanding transcendentality properties of Feynman integrals \cite{Henn:2013pwa}. For our purposes we will see that it allows us to ``exponentiate'' the action of loops $\gamma$, such as that in \eqref{fig:loop}, when transporting amplitudes on the kinematic space. The use of dimensional regularization---which was not introduced until 1970's when the S-matrix program was already on hiatus---is actually the key difference between our approach and that of \cite{FOTIADI1965159,AnalyticStudy1,AnalyticStudy2}. Had it not have been for the unfortunate history of the subject, we have no doubt the results discussed in this article would have been discovered much sooner.

It will be useful to keep an explicit example on the back of our heads. Throughout this article we will focus on the arguably simplest class of four-point one-loop diagrams with no masses: 
\be\label{box-family}
I_{n_1 n_2 n_3 n_4}(s,t) = \int_\Gamma \frac{d^{4-2\varepsilon} \ell}{[\ell^2]^{n_1} [(\ell{+}p_1)^2]^{n_2} [(\ell{+}p_1{+}p_2)^2]^{n_3} [(\ell{+}p_4)^2]^{n_4} },
\ee
which only depends on $s,t$ and $\varepsilon$.
The integration contour $\Gamma$ is chosen to impose the correct causality conditions. Here $n_a$'s are integers that indicate whether a given propagator is present in a specific diagram or not. For instance, the box, triangle, and bubble diagrams can be written as
\be\label{diagrams}
\includegraphics[scale=1,valign=c]{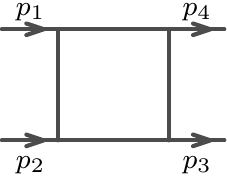} = I_{1111}, \qquad \includegraphics[scale=1,valign=c]{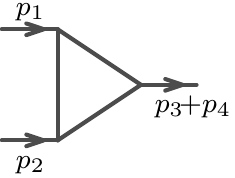} =  I_{1110}, \qquad \includegraphics[scale=1,valign=c]{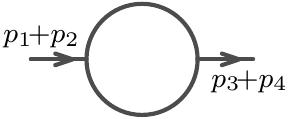} = I_{1010}.
\ee
We also allow for a possibility of integers other than $0$ and $1$, which arise in many contexts, such as differential equations studied below.

We can now return to the question of parallel transport on the kinematic space. Starting with some Feynman integral, say the box $I_{1111}$, computed at a given point $(s_\ast,t_\ast)$ in the kinematic space, let us transport it along $\gamma$ and \emph{demand} that the answer can be written as an exponential $e^{\int_{\gamma} \!\phi} I_{1111}$ for some integrand $\phi$ for all $\gamma$'s (for now this requirement might seem ad hoc, but it will become clear why it was made later in the text). If this is not the case, then we need to enlarge our original ansatz and consider a \emph{vector} of two integrals, say $(I_{1111}, I_{1110})^\intercal$, and ask whether for any $\gamma$ it can be written as a path-ordered exponential ${\cal P} e^{ \int_\gamma \!\bm{\phi}} (I_{1111}, I_{1110})^\intercal$ with some $2{\times} 2$ matrix $\bm{\phi}$. If not then we need to enlarge the ansatz once again, and so on. It is not terribly obvious that this process should truncate after a finite number of steps, but one can prove that it does. As a matter of fact, this number turns out to be a topological invariant of the internal loop momentum space! Let us denote it by $\chi$.

For instance, for the specific case of the family of integrals \eqref{box-family} this process turns out to truncate after three steps, so $\chi=3$ (let us mention that this number will in general be different for other loop orders). Let us call the vector of integrals at the starting point $(s_\ast,t_\ast)$ by $|\Phi\ra$, e.g.,
\be\label{intro-basis}
|\Phi \ra = \big( I_{1111},\, I_{1110},\, I_{1010} \big)^\intercal
\ee
and each element $| \Phi_i \ra$ for $i=1,2,\ldots, \chi$. Right now the notation with the ``ket'' vector $|\Phi\ra$ might seem silly, but it will become clear shortly why we decided to use it. Please do not confuse it for a quantum state. To abstract away from this specific example, let us denote the coordinates on the kinematic space by $x^\mu$ from now on. For instance, here $(x^1, x^2) = (s,t)$.

We may think of the vector $| \Phi \ra$ as belonging to some vector space $\V_{x}$ attached to a given point $x$ in the kinematic space. Clearly, for any other point $x'$ we can define an isomorphic vector space $\V_{x'}$ with its own vectors $|\Phi'\ra$. As a cartoon, let us keep in mind the following picture:
\be
\includegraphics[scale=1,valign=c]{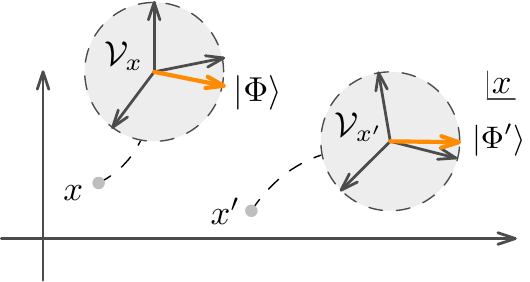}
\ee
These spaces can be ``glued'' together by parallel transport into a structure called \emph{vector bundle}. A physicists should think of it as a non-Abelian gauge field on the kinematic space. So far we do not know much about it, except that its gauge group is $\GL(\chi)$ (the space of $\chi {\times} \chi$ matrices). We would like to find out more.

As usual, we approach the problem infinitesimally: consider a point $x' = x + \delta x$ for small displacement $\delta x$. The vector $|\Phi' \ra$ and be written as $| \Phi' \ra = | \Phi \ra + \delta |\Phi \ra$, where
\be\label{perturbation}
\delta|\Phi \ra = \partial_{\mu}|\Phi \ra \,\delta x^\mu.
\ee
We make use of the summation convention for repeated indices. In the infinitesimal limit $\delta x^\mu \to 0$ we can treat the perturbation $\partial_{\mu} | \Phi \ra$ as living in the same vector space $\V_x$ as $|\Phi\ra$, which means we can express the former in terms of the latter. Just as in a linear algebra class, in order to project vectors onto each other we need a notion of a scalar product. This in turn requires us to introduce a \emph{dual} vector space $\V^{\vee}_x$ to $\V_x$. It will turn out that $\V^\vee_x$ is the vector space of Feynman integrals defined in $4{+}2\varepsilon$ instead of $4{-}2\varepsilon$ space-time dimensions, however we will not make use of this fact until later on in the article. For the time being we can just assume that \emph{some} dual space $\V^\vee_x$ exists and keep the discussion more abstract at first, before moving on to its specific implementation further down the line.

Let us also assume there exists a scalar product between orthonormal basis vectors $\la \Phi^\vee | \in \V_x^\vee$ and $| \Phi \ra \in \V_x$, i.e.,
\be\label{orthonormality}
\la \Phi^\vee_i | \Phi_j \ra = \delta_{ij}.
\ee
In other words, we have the resolution of identity:
\be
\mathds{1} = | \Phi_i \ra  \la \Phi^\vee_i |,
\ee
which can be confirmed by contracting the above expression from the left by $\la \Phi_j^\vee |$, from the right by $| \Phi_k \ra$, and using \eqref{orthonormality}. With these results in place we can go back to the problem of expressing $\partial_\mu |\Phi \ra$ in terms of $|\Phi \ra$. Inserting a resolution of identity we find
\be
\partial_{\mu} | \Phi_i \ra = | \Phi_j \ra \la \Phi^\vee_j | \partial_{\mu} | \Phi_i \ra.
\ee
Hence the transformation between the vectors $\partial_{\mu} |\Phi\ra$ and $|\Phi\ra$ of our interest is really given by the scalar products above. Let us organize them into a matrix $\mathbf{\Omega}_\mu$ with entries given by $(\mathbf{\Omega}_{\mu})_{ij} = \la \Phi^\vee_j | \partial_{\mu} | \Phi_i \ra$. This hints that we should define a covariant derivative $D_\mu = \partial_{\mu} - \mathbf{\Omega}_\mu$, which gives the ``equation of motion'' for $|\Phi\ra$:
\be\label{intro-differential}
D_\mu |\Phi \ra = 0,
\ee
or with indices
\be
D_\mu | \Phi_i \ra = \partial_\mu |\Phi_i\ra - (\mathbf{\Omega}_{\mu})_{ij} |\Phi_j\ra = 0.
\ee
In the field-theoretic language, what we found is that $\mathbf{\Omega}_\mu$ is a gauge field valued in $\GL(\chi)$ and
vectors of Feynman integrals transform as minimally-coupled scalars $|\Phi\ra$ in the fundamental representation. Gauge transformations are then
\be
\mathbf{\Omega}_\mu \;\mapsto\; \mathbf{g}\, \mathbf{\Omega}_{\mu}\, \mathbf{g}^{-1} + \partial_{\mu} \mathbf{g}\, \mathbf{g}^{-1} , \qquad |\Phi\ra \;\mapsto\; \mathbf{g}\, |\Phi\ra,
\ee
for any element $\mathbf{g} \in \GL(\chi)$. It means that the choice of the basis Feynman integrals, such as the one we made in \eqref{intro-basis}, is just a gauge choice. As usual, there might be good or bad choices of gauge and we will discuss which are the convenient ones later in the text.

When talking about the curvature of the gauge field, given by $\mathbf{F}_{\mu\nu} = -[D_\mu, D_\nu]$, there are two equivalent points of view. One way is to excise the singular points (e.g., positions of thresholds) from the kinematic space, which makes it a topologically non-trivial manifold. In this case the gauge field $\mathbf{\Omega}_\mu$ is always flat (integrable), i.e.,
\be
\mathbf{F}_{\mu\nu} = \partial_\mu \mathbf{\Omega}_\nu - \partial_\nu \mathbf{\Omega}_\mu - [\mathbf{\Omega}_\mu, \mathbf{\Omega}_\nu] = 0,
\ee
and hence locally it can be written as a pure gauge $\mathbf{\Omega}_\mu = \partial_{\mu} \mathbf{\Lambda}\, \mathbf{\Lambda}^{-1}$ for some matrix $\mathbf{\Lambda}$.
Another point of view is to keep the singular points as a part of the kinematic space, in which case they act as source currents for the gauge field. In any case, the paths $\gamma$ described at the beginning give Wilson loops
\be
\mathcal{P} \exp \int_{\gamma} \mathbf{\Omega}_\mu dx^\mu,
\ee
which are non-trivial either for topological reasons, or because they enclose a source, in the two interpretations respectively. Clearly, $\mathbf{\Omega}_{\mu}$ knows everything about the analytic structure on the kinematic space. For example, simple poles of $\mathbf{\Omega}_{\mu}$ determine positions of branch points and residues around these poles compute the discontinuities.

To make the above picture complete we need to be able to find out how to define more concretely: (a) the vector space $\V_x$, (b) the dual vector space $\V^\vee_x$, and crucially (c) how to define and compute the scalar product between them. These will be addressed in turn in the following sections. The points (a) and (b) will have a simple solution, which roughly speaking corresponds to a similar gauge-field structure, but on the internal loop-momentum space instead of the external kinematics space.

The scalar product between Feynman integrals turns out to be a new class of geometric invariants, which despite being studied by mathematicians for decades \cite{cho1995,zbMATH03996010,saito1983higher}, appeared in physics only recently \cite{Mizera:2017rqa,Mastrolia:2018uzb}. They are called \emph{intersection numbers}. If $\la \varphi_- |$ represents a specific diagram in $4{+}2\varepsilon$ dimensions, and $| \varphi_+ \ra$ is one in $4{-}2\varepsilon$ dimensions, we write their intersection number as $\la \varphi_- | \varphi_+ \ra$.
What is remarkable is that for an arbitrarily complicated Feynman integral, with any number of loops and legs, massive or massless propagators, planar or non-planar, the result is always a rational function of kinematic invariants and $\varepsilon$! This fact is in stark contrast with Feynman integrals themselves, which in principle involve as complicated functions as allowed by algebraic geometry.

Intersection numbers are of both theoretical and practical significance. We have already seen that they govern the differential equations on the kinematic space through the connection matrix $\mathbf{\Omega}_\mu$. They can be also used to project an arbitrary Feynman integral into a basis of integrals in the same topology. For instance, for a one-loop matrix element ${\cal T}_{12 \to 34}^{\text{one-loop}}$, the expansion in terms of the box, triangle, and bubble coefficients is given by:
\be
{\cal T}_{12 \to 34}^{\text{one-loop}}  = I_{1111}\, \la I_{1111}^\vee | {\cal T}_{12 \to 34}^{\text{one-loop}} \ra + I_{1110}\, \la I_{1110}^\vee | {\cal T}_{12 \to 34}^{\text{one-loop}} \ra +  I_{1010}\, \la I_{1010}^\vee | {\cal T}_{12 \to 34}^{\text{one-loop}} \ra,
\ee
where the coefficients are given by intersection numbers. We will also see that this formalism reveals a surprise about Feynman integrals: even though we are interested in the limit $\varepsilon \to 0$, which corresponds to four-dimensional physics, under some circumstances it will be possible to extract exactly the same information from the opposite limit, $\varepsilon \to \infty$ \cite{Mizera:2019vvs}!

This article is meant as an exposition aimed at physicists interested in understanding the broad-strokes picture behind the above ideas. We will attempt to avoid doing technical computations or using sophisticated mathematics. Most of the results discussed here are based on the recent papers \cite{Mizera:2017rqa,Mastrolia:2018uzb,Frellesvig:2019kgj,Mizera:2019gea,Frellesvig:2019uqt,Mizera:2019vvs}, where one can find more detailed discussion. In the final section and throughout the article we give a list of open questions, which the reader is invited to consider.

\section{\label{sec:Feynman}Feynman Integrals}

We begin by reviewing the definition of Feynman integrals in dimensional regularization. As with any other function, the way it is written as an integral is not unique. Over the years different representations have been introduced, each suitable for its own purpose. Most of them take the general form:
\be\label{integral}
I_{i} = \int_{\Gamma} e^{\varepsilon W} \varphi_i.
\ee
Here $W$ is in general a multi-valued function and $\Gamma$ is an integration domain, both of which are common to all Feynman diagrams in a given family. Let us assume that causality conditions are already implemented in the choice of $\Gamma$, so we do not have to worry about the $i\epsilon$ prescription. We will call $W$ a \emph{potential} for reasons that will become clear later in the text. The point of writing Feynman integrals as in \eqref{integral} is to emphasize the separation between the universal objects ($W$ and $\Gamma$) and the ones associated to a choice of a specific diagram ($\varphi_i$). Throughout this paper we will use the language of differential forms (for introduction, see, e.g., \cite[Ch.~5]{Nakahara:2003nw}), which will greatly simplify the notation. For example, $\varphi_i$ is a top (maximal degree) holomorphic form.

Let us review different representations, which differ by the physical meaning of the potential $W$ and a prescription for choosing $\varphi_i$. Since most of the readers should be familiar with them, and this part is not essential to understanding the ideas of the rest of the paper, we will be brief and partly schematic.

\paragraph{Loop Momentum Representation.} This is the most common way of writing Feynman integrals, which we already used in \eqref{box-family}. Isolating contributions coming from the  $-2\varepsilon$ dimension is precisely what gives a non-trivial potential $W$, which schematically takes the form
\be
W = -\log (\text{Lorentz invariants of momenta in the } {-}2\varepsilon \text{ dimensions}).
\ee
When the number of loops is $L$, the total number of integration variables becomes $4L + L(L+1)/2$ including the extra dimensions. For instance, for the box diagram \eqref{box-family} the integration space is given by the components of the loop momentum $(\ell^0, \ell^1, \ell^2, \ell^3, \ell^\perp)$, where the final one lives in the ${-}2\varepsilon$ dimensions. The potential in this case is simply
\be\label{W-perp}
W = - \log( (\ell^\perp)^2 )
\ee
and $\varphi$'s are proportional to $d\ell^0 \w d\ell^1 \w d\ell^2 \w d\ell^3 \w d\ell^\perp$ with appropriate denominators. For details see, e.g., \cite{CHP}. In the present context, this representation will most likely be the most useful for the study of full scattering amplitudes at a given loop order (not just individual families of scalar integrals). It is also related to the theory of hypersphere arrangements \cite{AIF_2003__53_4_977_0,Aomoto:2017npl}.

\paragraph{Baikov Representation.} A convenient way of performing loop integration is to first translate the loop momenta $\ell_i^\mu$ into independent Lorentz invariants $\ell_i {\cdot} p_j$ and $\ell_i {\cdot} \ell_j$. For $n$-point scattering this results in $L \min(n{-}1,4) + L(L{+}1)/2$ integration variables. Jacobian for this change of variables gives the so-called Baikov polynomial \cite{Baikov:1996iu}, which enters the potential $W$:
\be
W = - \log(\text{Baikov polynomial}).
\ee
We refer to \cite[App.~A]{Mastrolia:2018uzb} for details. This representation has a subtlety in that only for diagrams with $\min(n{-}1,4){+}L$ odd it takes the specific form \eqref{integral} (when this combination is even we need to use $e^{\varepsilon W} \to e^{(\varepsilon{-}\frac{1}{2})W}$ instead), and hence some of the later discussion would have to be modified to capture those cases. Nonetheless, it becomes very useful for studying cuts \cite{Frellesvig:2017aai}: vanishing of a single propagator defines a hyperplane in the integration space (as opposed to a quadric in the original loop momentum variables) and computing a cut corresponds to taking a residue around it. In the box example \eqref{box-family} the integration variables can be taken to be $(\ell^2, \ell{\cdot}p_1, \ell{\cdot}p_2, \ell{\cdot}p_3)$, which results in
\be
W = - \log \Big(\ell^2 - \sum_{a,b=1}^{3} \ell{\cdot} p_a\, \mathbf{G}_{ab}^{-1}\, p_b {\cdot} \ell \Big),
\ee
where $\mathbf{G}$ is the Gram matrix with entries $\mathbf{G}_{ab} = p_a {\cdot} p_b$. This is of course just another way of writing \eqref{W-perp}, since $\ell^2 = (\ell^\perp)^2 + (\ell^{\parallel})^2$ and the four-dimensional norm $(\ell^\parallel)^2$ can be projected onto Lorentz invariants using the sum above.

\paragraph{Feynman Representation.} Here we trade loop variables for Schwinger parameters, which measure proper time $z_a$ along the $a$-th propagator. The resulting integral is commonly expressed in terms of the so-called Symanzik polynomials ${\cal F}$ and ${\cal U}$, which are respectively degree $L{+}1$ and $L$ polynomials in $z_a$'s, see, e.g., \cite{Bogner:2010kv} for a review. We will consider a version of this parametrization popularized by \cite{Lee:2013hzt}, which results in
\be\label{W}
W = \log ({\cal F} + {\cal U})
\ee
and the forms $\varphi_i$ are now labeled by the integers $(n_1, n_2, \ldots, n_\P)$ for each of the $\P$ propagators, 
\be\label{twisted-forms}
\varphi_{n_1 n_2 \cdots n_\P} = \frac{1}{({\cal F}{+}{\cal U})^2} \bigwedge_{a=1}^{\P} \frac{dz_a}{z_a^{1- n_a}}.
\ee
There is a kinematics-dependent overall constant that we ignore for the purposes of our discussion, see, e.g., \cite[Sec~3.1]{Mizera:2019vvs} for the full derivation. We will refer to this form of Feynman integrals as the \emph{Feynman representation} (this terminology is not standard). As in the worldline formalism, physically $z_a$'s are the proper times along a given edge of the Feynman diagram. Each of them ranges from $0$ to $\infty$, and hence the domain of the integration is $\Gamma = \R_+^\P$. In the box example \eqref{box-family} we have $\P{=}4$ and find:
\be\label{FU}
{\cal F} = s z_1 z_3 + t z_2 z_4, \qquad {\cal U} = z_1 + z_2 + z_3 + z_4.
\ee
The three diagrams in \eqref{diagrams} are computed with
\be
\varphi_{1111} =  \frac{1}{({\cal F}{+}{\cal U})^2} d^4z, \qquad \varphi_{1110} =  \frac{1}{({\cal F}{+}{\cal U})^2} \frac{d^4z}{z_4}, \qquad \varphi_{1010} =  \frac{1}{({\cal F}{+}{\cal U})^2} \frac{d^4z}{z_2 z_4},
\ee
where $d^4 z = dz_1 \w dz_2 \w dz_3 \w dz_4$.

\vspace{1em}
We will use the Feynman representation in the examples throughout the article, though it is important to remember that all the discussion can be repeated for other representations with essentially no changes, which was the reason for unifying them in the common expression \eqref{integral} in the first place.

\subsection{Types of Boundaries}

In order talk more precisely about the geometry of integrals such as \eqref{integral}, we need to define the space on which they are computed. Let us call this complex manifold $M$ and its complex dimension $\P$ (the real dimension is $2\P$).

In the Feynman representation it is convenient to treat each Schwinger parameter $z_a$ as an inhomogeneous variable on a Riemann sphere $\CP^1$ (this is just a complex plane with a point at infinity included, $\CP^1 = \C \cup \{\infty\}$), even though it is integrated only along the real positive half-line. To define $M$ we have to remove places where the integrand $e^{\varepsilon W} \varphi_i$ can diverge, which are determined by ${\cal F} {+} {\cal U} = 0$ and by $z_a = 0,\infty$ for all $a$. We will refer to those places as \emph{boundaries}. Therefore the space $M$ is given by $\P$ copies of $\C^\ast = \CP^1 {-} \{0,\infty\}$ with the aforementioned hypersurface excised:
\be
M = (\mathbb{C}^\ast)^\P - \{ {\cal F} {+} {\cal U} = 0\}.
\ee
Boundaries of $M$ are sources of ultraviolet and infrared divergences. Mathematically speaking they fall into one of the following two categories. 

\paragraph{Twisted Boundaries.} These are boundaries regulated by $\varepsilon$. They can happen when $e^{\varepsilon W}$ goes to zero or infinity. If we treat $\varepsilon$ as a generic number or a formal variable, then the integral integrates to a well-defined expression in the neighborhood of those boundaries. In our case $\{ {\cal F} {+} {\cal U} = 0\}$ defines a twisted boundary.

\paragraph{Relative Boundaries.} These are the unregulated boundaries. They happen when the form $\varphi_i$ has poles that are not affected by $e^{\varepsilon W}$. The integral can in principle diverge close to them. In our case relative boundaries are given by $\{z_a = 0,\infty\}$ for any $a$.

\vspace{1em}
Note that if we were working without dimensional regularization, in whatever representation of Feynman integrals, all boundaries would be relative. (In those cases another complication arises when the integrand contains square roots, as then we need to consider double covers of $M$. The simplest examples are elliptic curves. Avoiding such situations is one motivation for using dimensional regularization.)

As an example, let us consider Symanzik polynomials \eqref{FU} and draw a cartoon of the real cross-section of $M$ for some fixed values of $(z_2,z_4)$:
\be
\includegraphics[scale=1,valign=c]{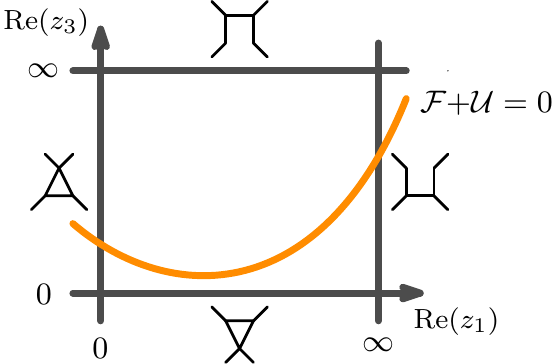}
\ee
The twisted boundary is illustrated in orange. The relative boundaries where the $z_a$ goes to either zero or infinity (gray) have a combinatorial interpretation as the so-called contraction-deletion relations, which correspond to propagators being either pinched or removed, see, e.g., \cite[Sec.~6]{Bogner:2010kv}.

Although it is not strictly necessary, from now on we will implement an extra simplification, which translates all relative boundaries into twisted ones. This is achieved by simply deforming our potential to
\be\label{W-regulated}
W = \log ({\cal F}{+}{\cal U}) + \sum_{a=1}^{\P} \delta_a \log z_a,
\ee
which by the above definitions makes $\{z_a = 0,\infty\}$ twisted. Here $\delta_a$'s are some additional parameters (one can set them all equal, $\delta_a {=} \delta$), which are sent to zero at the end of a computation. One can think of them as an additional regulator on top of dimensional regularization (see also \cite[Ch.~3]{10.2307/j.ctt1b9x0ss}). Equivalent way of thinking is that we deform all integers $n_a$ into generic non-integer parameters $n_a + \varepsilon \delta_a$. Let us stress that this is a step which is only taken to simplify our discussion and should not be thought of as particularly fundamental.

\subsection{\label{sec:connection}Connection to Mathematics}

After the regularization described above, Feynman integrals take the form of generalized Aomoto--Gelfand hypergeometric functions \cite{aomoto1973theoreme,Gelfand}.\footnote{To be more concrete, when $W$ is given as in \eqref{W}, the integral \eqref{integral-pairing} falls into the class of resonant Gelfand--Kapranov--Zelevinsky hypergeometric functions \cite{GELFAND1990255,delaCruz:2019skx,Klausen:2019hrg}. For a historical summary of treating Feynman integrals as hypergeometric functions see \cite[Intro.]{Klausen:2019hrg}.} This fact in itself should not be surprising, as virtually all functions appearing in quantum field theory are of hypergeometric type. The key idea is to understand them as pairings of two objects:
\be\label{integral-pairing}
\la \Gamma \!\otimes\! e^{\varepsilon W} | \varphi_i \ra = \int_{\Gamma} e^{\varepsilon W} \varphi_i.
\ee
Here $\la \Gamma \otimes e^{\varepsilon W} |$ is an element of a vector space of all possible integration cycles. This space is called a \emph{twisted homology} group. It consists of two pieces of information: the integration domain $\Gamma$ and the branch of the multi-valued function $e^{\varepsilon W}$ we choose to integrate on. It is required that boundaries of $\Gamma$ lie along the twisted boundaries of $M$ discussed before. We will not discuss integration domains in more detail, since in our application they are always fixed (and the branches are uniquely specified by requiring that the integrand is real along $\Gamma$). 

Similarly, $\varphi_i$ denotes a single-valued differential $m$-form which can have poles only on the twisted boundaries. It is an element of another vector space of all possible integrands called \emph{twisted cohomology} group. We can understand it in elementary terms in the following way. In the absence of boundary terms, integral of any total differential vanishes, and so we have
\be\label{total-differential}
0 = \int_{\Gamma} d(e^{\varepsilon W} \xi) = \int_{\Gamma} e^{\varepsilon W} (d\xi + \varepsilon dW {\wedge} \xi)
\ee
for any $(m{-}1)$-form $\xi$ with poles only on the twisted (regulated) boundaries. Since the above combination will appear many times, let us introduce the notation $\nab = d + \varepsilon dW\wedge$, which is a covariant derivative.\footnote{In the literature the notation $\omega$ is often used instead of $dW$ for the same object, but we prefer to use the latter to make it manifest that $dW$ is a closed form. From there we immediately see that the covariant derivative is flat, $\nab^2 = ddW=0$.} The above result means that we can freely add combinations of the form $\nab \xi$ to $\varphi_i$ and it does not change the value of the integral \eqref{integral-pairing}. The integrand is therefore better thought of as an equivalence class $| \varphi_i \ra$ obtained by identifying
\be\label{equivalence}
\varphi_i \;\sim\; \varphi_i + \nab \xi
\ee
for any $\xi$. These classes form a finite-dimensional vector space known as the $\P$-th twisted cohomology group, denoted by $H^\P ((\C^\ast)^\P {-} \{{\cal F}{+}{\cal U}{=}0\}, \nab)$ in the mathematics literature, which for short we will call $H^\P_{\dW}$ from now on. Thus we have
\be
| \varphi_i \ra \;\in\; H^\P_{\dW}.
\ee
In mathematics such equivalence classes are called twisted cocycles. In physics we mostly deal with \emph{representatives} of these classes, such as the ones given in \eqref{twisted-forms}. We will refer to such representatives as \emph{twisted forms}, to distinguish them from ordinary differential forms.

This structure naturally parallels the one from Sec.~\ref{sec:introduction}. The covariant derivative $\nab$ is that associated to a flat Abelian gauge field with the gauge potential $dW$ valued in $\GL(1) = \C^\ast$. In this interpretation $\varphi_i$ can be identified as a $\P$-form field with gauge equivalence \eqref{equivalence}, though we do not suppose this interpretation gives any intuition in our application.

The key point is that for Feynman integrals expressed as the pairing \eqref{integral-pairing} the integration domain is always fixed. This means that choosing a specific integral $I_i$ is the same as choosing a twisted form $\varphi_i$. For example, in the Feynman representation we have
\be
I_{n_1 n_2 \cdots n_\P}(x) \;=\;  \la \underbrace{\R_+^\P \otimes e^{\varepsilon W(x)}}_{\text{fixed}} | \varphi_{n_1 n_2 \cdots n_\P}(x) \ra.
\ee
Here we temporarily reinstated the dependence on a point $x$ in the kinematic space. This motivates the identification of the vector space $\V_x$ of Feynman integrals as being modeled by the twisted cohomology group \cite{Frellesvig:2019uqt}:
\be\label{V-H}
\V_x \;\cong\; H_{\dW(x)}^{\P},
\ee
and hence the vectors $| \Phi_i \ra$ from Sec.~\ref{sec:introduction} are identified as $| \varphi_i \ra$.
This relationship is valid only up to the kernel of integration, which in reality is slightly bigger than that implied by \eqref{total-differential}. This is because $\la \R_+^\P \otimes e^{\varepsilon W} |$ might have additional symmetries. For instance, in the case of the box topology it is invariant under $z_1 \leftrightarrow z_3$ and separately under $z_2 \leftrightarrow z_4$, so for example $I_{1100} = I_{0011}$ on the level of the integrated answer, but $\varphi_{1100} \not\sim \varphi_{0011}$ on the level of cohomology. Another example is discussed in \cite[Sec.~3.4]{Mizera:2019vvs} for massive sunrise graph topology. In conclusion, the above construction is blind to such symmetry relations, though it seems likely that it can be modified to account for it.

The introduction of $\delta$-regulators was necessary to formulate our problem in terms of a twisted cohomology. In the presence of relative boundaries one needs to consider a richer structure called \emph{relative twisted cohomology}, which is the proper way of understanding Feynman integrals in dimensional regularization, but falls beyond the scope of this review. The simplification of not having relative boundaries around will allows us to invoke some of the results from mathematics without further complications, thus streamlining the discussion, but should not be regarded as the final word.

At this stage let us emphasize that even after employing $\varepsilon$- and $\delta$-regularization, the integrals \eqref{integral-pairing} are still \emph{not} convergent in general. By this we mean that one could not simply plug in any numerical values for kinematic variables and $\varepsilon$ and expect numerical integration on a computer to converge. This is actually a generic feature of integrals on non-compact spaces, such as $M$ (recall that any manifold with a boundary is non-compact). A formal way of dealing with this problem is to find a differential form $\varphi^c_i$, which is in the same class as the original $\varphi_i$,
\be
\varphi^c_i = \varphi_i + \nab \Xi
\ee
for some $\Xi$, in such a way that $\varphi^c_i$ vanishes in the small neighborhood of every boundary. A form with this property is said to have \emph{compact support}. In this way, the integrand is always bounded and the corresponding integral
\be\label{regularization}
\int_{\Gamma} e^{\varepsilon W} \varphi_i^c
\ee
converges. Strictly speaking, in the definition \eqref{integral-pairing} we should have used this version. Even though it can be proven that $\varphi^c_i$ exists for the above class of integrals \cite[Sec.~2.2]{aomoto2011theory}, it can be difficult to find it in practice (a more pragmatic approach to numerical integration is sector decomposition, see, e.g., \cite{Bogner:2007cr}). We will return to this point in Sec.~\ref{sec:intersection} when discussing intersection numbers.

It might not be immediately clear why the above mathematical formulation would help us with anything, and indeed one can argue that so far it has not. The advantage we gained, however, is that we can use tools of algebraic topology and geometry to ask physics questions. The first one is: how many linearly-independent integrals of the above form are there?

\subsection{\label{sec:more}More Connections to Mathematics}

The above question formulated in the geometric language asks about the dimension of $H^\P_{\dW}$. Actually, one can attempt to construct other twisted cohomology groups $H^k_{\dW}$, which are spaces of $k$-forms (where $k$ is not necessarily equal to $\P$) up to equivalence relations \eqref{equivalence}, and ask a similar question.
Here we can refer to an important result of Aomoto \cite{aomoto1975vanishing}, who showed that all $H^{k}_{\dW}$ are empty unless $k = \P$ (this result holds under some genericity condition on $dW$, which are satisfied in our case). In other words, lower- and higher-degree forms with the transformation property \eqref{equivalence} do not exist. This statement becomes very powerful once we realize that dimensions of twisted cohomology groups defined above are related to the topological Euler characteristic $\chi(M)$ via
\be
\chi(M) = \sum_{k=0}^{2\P} (-1)^{k} \dim H_{\dW}^{k}.
\ee
Using the fact that $\dim H_{\dW}^{k\neq \P} = 0$ leaves us with
\be\label{dimension}
\dim H^\P_{\dW} = (-1)^k \chi(M),
\ee
so up to a sign the number of linearly-independent Feynman integrals computed on $M$ is given by $\chi(M)$. For a general introduction to Euler characteristics see \cite[Sec.~2.4.4]{Nakahara:2003nw}. In the present formulation this result was given in \cite{Mastrolia:2018uzb,Frellesvig:2019uqt,Mizera:2019vvs}, but can be also derived by other means \cite{Bitoun:2017nre}.

This result is particularly interesting, because it translates a physical problem into a geometric one. At this stage we can exploit the fact that there are multiple different ways of computing Euler characteristics. Particularly useful for us will be the connection with \emph{Morse theory}, which, broadly speaking, allows one to study topology of a manifold using flow equations of a ``height function'' defined on it. See \cite[Sec.~3]{Witten:2010cx} for an introduction.

In our case a natural choice of the height function is $\text{Re}(W)$. Taking the real part is quite important to make it single-valued and also to define the ``height'' associated to each point on $M$, which is supposed to be a real number. The height diverges to $\pm \infty$ at all the boundaries of $M$. It must then have some critical points in between (critical points are places where the first derivative of $\text{Re}(W)$ vanishes). To each of them we associate an \emph{index} $\gamma$, which counts how many downwards directions extend from this point. Since $M$ is a $2\P$-real-dimensional manifold, the index has to be between $0$ and $2\P$. Let us call the number of critical points with index $\gamma$ by $C_\gamma$. Morse theory tells us that the Euler characteristic can be expressed as
\be\label{Morse}
\chi(M) = \sum_{\gamma=0}^{2\P} (-1)^\gamma C_\gamma.
\ee
This sum has only one non-vanishing term, since one can show that $\text{Re}(W)$ for any holomorphic function $W$ has only critical points with indices $\gamma=\P$ (see, e.g., \cite[Sec.~2.4.3]{Mizera:2019gea}). This means that the height function near all critical points has a shape of saddle (with equal number of upwards and downwards directions extending from it). Consequently $C_{\gamma \neq \P} = 0$ and $C_{\P}$ counts the total number of critical points. Then using \eqref{Morse} together with our previous result \eqref{dimension}, we find
\be\label{dim-H}
\dim H^\P_{\dW} = \text{total number of critical points of }\text{Re}(W).
\ee
In addition, it is not difficult to see that positions of critical points of $\text{Re}(W)$ are the same as those of $W$, and hence they can be determined algebraically by solving
\be\label{dW}
dW= \sum_{a=1}^{\P} \frac{\partial W}{\partial z_a} dz_a =  0,
\ee
which asks for coefficients of each $dz_a$ to vanish, giving a system of $\P$ equations:
\be
\frac{\partial W}{\partial z_a} = 0, \qquad a=1,2,\ldots,\P.
\ee
The number of solutions is equal to $|\chi(M)|$, which for short we will call simply $\chi$ in the future.

For example, by plugging \eqref{FU} and \eqref{W-regulated} into the above equation, the number of independent Feynman integrals in the box graph topology is found by solving the system of equations
\be\label{box-dW}
\frac{\delta_1}{z_1} + \frac{1+s z_3}{{\cal F}+\cal{U}} =0, \qquad \frac{\delta_2 }{z_2} + \frac{1 + t z_4}{{\cal F}+\cal{U}}=0,\qquad \frac{\delta_3
}{z_3}+\frac{1+s z_1}{{\cal F}+\cal{U}}=0,\qquad \frac{\delta_4}{z_4}+\frac{1 + t z_2}{{\cal F}+\cal{U}}=0,
\ee
which has $\chi=3$ solutions, confirming the assertion made in Sec.~\ref{sec:introduction}. Note that since we only care about the number of solutions, but not their explicit form, we can do this computation numerically for random values of external parameters $(s,t,\delta_1,\delta_2,\delta_3,\delta_4)$, which turns it into a very efficient method of counting Feynman integrals.

Strictly speaking, in the equality \eqref{Morse} there was an assumption that all critical points are isolated (the are no continuous families) and non-degenerate (second derivatives of $W$ do not vanish). With small modifications one can relax these assumptions. For example, \eqref{dim-H} still holds if we take into account multiplicity of each degenerate critical point or alternatively perturb the height function so that the degenerate point splits into a number of non-degenerate ones infinitesimally far away from each other. For treating non-isolated critical points see \cite[Sec.~5]{Lee:2013hzt}.

Physically, it might seem rather surprising that critical points have something to say about counting Feynman integrals in dimensional regularization, since they are normally associated tho the saddle-point approximation in the $\varepsilon \to \infty$ limit, rather than the physical $\varepsilon \to 0$ that gets us to four dimensions. In Sec.~\ref{sec:epsilon} we will see an even more striking example of this phenomenon and understand it in terms of intersection theory.

As mentioned before, there are multiple other ways of computing Euler characteristics, each of which can serve as a proxy for determining the number of linearly-independent Feynman integrals. Other than the method mentioned above, one can use the computer package \texttt{Macaulay2} \cite{M2}, which computes Euler characteristics with computational algebraic geometry algorithms (see \cite[App.~A]{Cachazo:2019ngv} for an example usage). Let us also mention methods using the theory of Chern--Schwartz--MacPherson classes \cite{Aluffi:2008rw}, point counting over finite fields \cite{belkale2003,Brown:2010bw}, and volumes of Newton polytopes \cite{Bitoun:2017nre,Klausen:2019hrg}. Note that these techniques appeared previously only in the context of Feynman integrals \emph{without} dimensional regularization, where $M$ is typically given as a complement of $\{{\cal U} =0\}$ hypersurface instead of $\{{\cal F}{+}{\cal U}=0\}$ considered here.

At this stage it is worth to pause for a second and ask the following question.

\subsection{What Does Euler Characteristic Really Count?}

One convenient way of thinking about the dimension of the twisted cohomology group is as computing the rank of a matrix of integrals \eqref{integral} for all possible integration domains (labeled by rows) and integrands (labeled by columns), i.e.,
\be
\chi = \text{rank} \setlength\arraycolsep{.5em}\begin{pmatrix}
	\displaystyle\int_{\Gamma_1}\!\! e^{\varepsilon W} \varphi_1 & \displaystyle\int_{\Gamma_1}\!\! e^{\varepsilon W} \varphi_2 & \cdots\quad  \\[1em]
	\displaystyle\int_{\Gamma_2}\!\! e^{\varepsilon W} \varphi_1 & \displaystyle\int_{\Gamma_2}\!\! e^{\varepsilon W} \varphi_2 & \cdots\quad \\[1em]
	\vdots	& \vdots & \ddots\quad \\[.1em] 
\end{pmatrix}.
\ee
If we were interested in the number of independent integrals for \emph{all possible} choices of $\Gamma_i$ and $\varphi_j$ then the answer would be $\chi^2$. In our specific problem, however, we care only about the integrals for a \emph{specific} contour $\Gamma = \R_+^\P$, which corresponds to a single row of the above matrix. Number of linearly-independent integrals of this type is therefore at most $\chi$, but might in principle be lower. One reason for this was discussed underneath \eqref{V-H}.

It is also important to mention that strictly speaking in the above discussion we were talking about $\delta$-regulated Feynman integrals obtained by sending $n_a \to n_a + \varepsilon \delta_a$, whose counting might conceivably differ from the number of the original dimensionally-regularized Feynman integrals before this deformation. Of course, the number of independent integrals also jumps discontinuously as we approach four dimensions in the strict limit $\varepsilon =0$.

In the literature of Feynman integrals the term ``master integrals'' is often used to refer to a basis of Feynman integrals. Given its ambiguity we prefer not to use this term here. For example, to define a basis one needs to specify a coefficient field (what coefficients of a basis expansion are allowed to be). In our setup we work in the rational field of coordinates of the kinematic space and $\varepsilon$, for example $\Q(s,t,\varepsilon)$ in the box topology case (to explain this we need to wait until Sec.~\ref{sec:intersection}). In contrast, performing the counting of integrals over $\Q$ is likely to give a higher answer, while adding variables (for example, square roots) to the coefficient field might conceivably make it smaller. As an additional warning let us also mention that some authors might count Feynman integrals assuming: permutation symmetries, different treatment of subsectors, exclusion of reducible diagrams, or counting of only top-level integrals, among others (see also the discussion in \cite[Sec.~4.4]{Bitoun:2017nre}). One motivation for introducing twisted cohomology groups was to systematize the definition of the term ``master integrals''.

Lastly, let us comment on the relation to the work of Lee and Pomeransky \cite{Lee:2013hzt}. Translated to our notation they argued that the number of linearly-independent integrals in the top sector (on the maximal cut) is counted by the number of critical points of $\text{Re}(W(\delta_a {=}0))$, i.e., without the additional regulators. Crucially, the system of equations $dW=0$ determining critical points is in principle discontinuous when any $\delta_a \to 0$. For example, setting $\delta_a = 0$ in \eqref{box-dW} \emph{before} solving the equations yields the system
\be
\frac{1+s z_3}{{\cal F}+\cal{U}} =0, \qquad \frac{1 + t z_4}{{\cal F}+\cal{U}}=0,\qquad \frac{1+s z_1}{{\cal F}+\cal{U}}=0,\qquad \frac{1 + t z_2}{{\cal F}+\cal{U}}=0,
\ee
which has only $1$ solution. This is consistent with the fact that box topology has only one independent integral on the maximal cut (for example $I_{1111}$). In general there should be a way of predicting number of Feynman integrals at any level between the maximal cut and the full integral by setting various $\delta_a$ to zero and studying the behavior of critical points in such limits. This is an open question which deserves further work.

\section{\label{sec:intersection}Intersection Theory}

In the previous section we found that the vector space of Feynman integrals $\V$ can be modeled by the cohomology group $H^\P_{\dW}$. As anticipated in Sec.~\ref{sec:introduction}, we will define the dual vector space $\V^\vee$ by simply flipping the sign of the dimensional regulators from $4{-}2\varepsilon$ to $4{+}2\varepsilon$. Replacing $\varepsilon \to - \varepsilon$ in the discussion above allows us to identify $\V^\vee$ with $H^\P_{\mdW}$.  For example, the ``dual'' Feynman integrals are given by
\be
\la  \varphi_i | \Gamma \otimes e^{-\varepsilon W} \ra = \int_{\Gamma} e^{-\varepsilon W} \varphi_i.
\ee
Here we have intentionally used an opposite bra-ket notation to that in \eqref{integral-pairing}, to distinguish between $| \varphi_i \ra$, defined in \eqref{equivalence}, from $\la \varphi_i | \in H^\P_{\mdW}$, which is the equivalence class
\be
\varphi_i \;\sim\; \varphi_i + \nabla_{\mdW} \xi.
\ee
To further distinguish between the two distinct classes we will use the notation $\varphi_\pm$ for generic representatives (twisted forms) of $H^\P_{\pmdW}$.

What it means for a vector space to be dual to another one is that there exists a non-degenerate bilinear pairing between them. Hence we would like to write down a bilinear of $\la \varphi_- |$ and $| \varphi_+ \ra$ that respects all cohomology relations. If this was possible, it would define a ``scalar product between Feynman integrals''. The most naive guess for such an object is
\be\label{intersection-guess}
\la \varphi_- | \varphi_+ \ra \stackrel{?}{=} \int_{M} \left( e^{-\varepsilon W} \varphi_- \right) \wedge \left( e^{+\varepsilon W} \varphi_+ \right),
\ee
which actually turns out to be quite close to the correct answer and needs only a small refinement.

\subsection{Scalar Product Between Feynman Integrals}

It is instructive to understand what goes right and what goes wrong with the formula \eqref{intersection-guess}. First of all, it is obviously a bilinear, i.e., for constants $\alpha,\beta$ it satisfies
\be
\la \varphi_- | \alpha \varphi_{+} + \beta \widetilde{\varphi}_{+} \ra \,=\, \alpha \la \varphi_{-} | \varphi_+ \ra \,+\, \beta \la \varphi_{-} | \widetilde{\varphi}_+ \ra,
\ee
and similarly for the linearity for the other twisted form $\varphi_-$. Secondly, we would like to show that it respects the equivalence relations for both twisted forms, which can be checked by confirming that $\la \nabla_{\mdW} \xi | \varphi_+\ra$ vanishes for every $\xi$ and $\varphi_+$ (and similarly for $\la \varphi_- | \nab \xi \ra$). Naively doing this computation we obtain
\be\label{dW-xi}
\la \nabla_{\mdW} \xi | \varphi_+\ra \;=\; \int_M d\left( e^{-\varepsilon W} \xi \right) \wedge \left( e^{+\varepsilon W} \varphi_+ \right) \;=\; \int_M d \left[ \left( e^{-\varepsilon W} \xi \right) \wedge \left( e^{+\varepsilon W} \varphi_+ \right) \right] ,
\ee
where in the second equality we used the fact that $d( e^{+\varepsilon W}\varphi_+) = 0$ for a top holomorphic form. What remains to argue is that the integral of a total derivative on the right-hand side vanishes. By Gauss theorem $\int_M d(\cdots) = \int_{\partial M} (\cdots)$, but in our case the integrand has singularities on the boundary $\partial M$, so the expression seems ill-defined.

This is actually a symptom of an earlier problem. Notice that in \eqref{intersection-guess} both $\varphi_{\pm}$ are top holomorphic forms, for which $\varphi_- \wedge \varphi_+ = 0$ identically. On the other hand, the integral seems to diverge close to the boundaries, which gives rise to a ``0/0 problem'' near $\partial M$. The way to regularize it is similar to what we have already seen around \eqref{regularization}, where we introduced a compactly-supported form $\varphi^c_+$ in the same cohomology class as $\varphi_+$, but which vanishes in the small neighborhood of each boundary (it does not really matter if we choose to regularize $\varphi_-$ or $\varphi_+$). This gives us the proper definition of the scalar product we were looking for \cite{Mastrolia:2018uzb}:
\be\label{intersection-number}
\la \varphi_- | \varphi_+ \ra_{\dW} = \left( \frac{-\varepsilon}{2\pi i}\right)^{\!\!\P} \int_M \varphi_- \wedge \varphi_+^c.
\ee
It is called the \emph{intersection number} of twisted forms $\varphi_-$ and $\varphi_+$ \cite{cho1995}.
Note that compared to \eqref{intersection-guess} we canceled the factors of $e^{\pm \varepsilon W}$. The expression still depends on $dW$ because of the compact support imposed on $\varphi_+^c$. To remember this fact we introduced the subscript ${}_{\dW}$ on the left-hand side. We also normalized the whole expressions for later convenience (recall that $\P$ is the complex dimension of $M$).

The regularization we introduced makes the right-hand side of \eqref{dW-xi} vanish and hence defines a bilinear between the cohomology classes $\la \varphi_- |$ and $| \varphi_+ \ra$. One can show that it is a non-degenerate pairing, meaning that a $\chi \times \chi$ matrix of intersection numbers between basis elements of both cohomologies has full rank \cite{cho1995}.

Even though the intersection number \eqref{intersection-number} is written as an integral, it is not an integral in the conventional sense. Since we performed our regularization only near the boundaries, it is still true that $\varphi_- \wedge \varphi_+^c = 0$ away from them. This is a sign of \emph{localization}: the above integral receives contributions only from small regions of $M$, which are the neighborhoods of each boundary. This signals that the result of such an integration must be much simpler than that of a full-blown integral. Indeed, we will see that \eqref{intersection-number} can be written in different ways as a sum of residues. The result always turns out to be a rational function of kinematic invariants and $\varepsilon$.\footnote{Alternatively we could have chosen the dual vector space to be anti-holomorphic such that the scalar product becomes $\int_M e^{\varepsilon \overbar{W}} \overbar{\varphi_i} \wedge e^{\varepsilon W} \varphi_j$, which would be a perfectly good definition, but it would lead to much harder computations since the integral does not localize.}

This is a good point to comment on the nomenclature. The words ``intersection number'' or ``intersection pairing'' are commonly used for all the different types of homology-homology and cohomology-cohomology bilinears, not necessarily those of twisted cohomology used here. For example, \cite{hwa1966homology} talks about intersection numbers (or ``Kronecker indices'') of relative homology groups, which are integers counting how many times contours intersect. Another example are intersection numbers encountered in the study of Feynman integrals in tropical mirror symmetry \cite{boehm2018tropical}. These are \emph{not} the same as the intersection numbers discussed in this article and should not be confused with them. In principle, in the context of Feynman integrals one can also talk about intersection theory for cycles $\la \Gamma \otimes e^{\varepsilon W}|$ (see, e.g., \cite[App.~A.2]{Mizera:2019gea}), but we will not review it here.\footnote{\label{sec:aside}%
	As a resource for the reader, let us briefly outline the history of intersection numbers of twisted cohomologies in mathematics and their applications to scattering amplitudes. The idea of twisted co/homology groups is quite an old one and dates back at least to Reidemeister \cite{reidemeister1935uberdeckungen} and Steenrod \cite{10.2307/1969099}. It was later applied to the theory of hypergeometric functions by Aomoto \cite{aomoto1973theoreme} and independently by Gelfand \cite{Gelfand}, which included the study of differential equations and linear relations between integrals. For a comprehensive introduction and a list of references see \cite{aomoto2011theory}. In order to study quadratic relations between hypergeometric integrals, in 1995 Cho and Matsumoto introduced intersection numbers of twisted cohomology classes \cite{cho1995}. Equivalent definitions can be found even in the earlier literature, most notably in the work of Deligne and Mostow \cite{zbMATH03996010} and Saito \cite{saito1983higher} in the 1980's, though the focus of these works was less practical. Later mathematical literature on computing intersection numbers in various contexts includes \cite{matsumoto1998,majima2000,OST2003,Yoshiaki-GOTO2015203,matsubaraheo2019euler,matsubaraheo2019algorithm}.
	
	In 2017 it was found that intersection numbers have a physical interpretation in terms of scattering amplitudes \cite{Mizera:2017rqa}. In this context, which parallels the developments described here, intersection numbers on the moduli space of genus-zero Riemann surfaces with $n$ punctures ${\cal M}_{0,n}$ compute $n$-point tree-level scattering amplitudes of different quantum field theories in a way alternative to Feynman diagrams. Together with their homological counterpart, they can be used to understand linear \cite{Mizera:2017cqs,Mizera:2019gea,Casali:2019ihm} and quadratic \cite{Mizera:2017cqs} relations between such scattering amplitudes, as well as their connections to color-kinematics duality \cite{Mizera:2019blq} and string theory \cite{Mizera:2017rqa,Mizera:2019gea}. Most likely, these computations can be extended to genus-$g$ moduli spaces ${\cal M}_{g,n}$, where intersection numbers are expected to compute $g$-loop integrands for different quantum field theories.
	
	With Mastrolia we later applied similar techniques to Feynman integrals \cite{Mastrolia:2018uzb}, where intersection numbers turn out to have another physical interpretation, which is the main focus of this review. We will return to possible connections between the two types of intersection numbers in Sec.~\ref{sec:summary}.
}

Before discussing how to compute intersection numbers, let us recall why we needed them. The first application comes from expanding an arbitrary Feynman integral in a basis. In order to achieve this, let us introduce a set of bases orthonormal with respect to the intersection pairing:
\be
\la \varphi^\vee_i | \varphi_j \ra_{\dW} = \delta_{ij}.
\ee
We will use this set to project integrals using the resolution of identity
\be\label{resolution}
\mathds{1} = | \varphi_i \ra \la \varphi_i^\vee |.
\ee
Recall that we use implicit summation convention for repeated Latin indices $i,j=1,2,\ldots,\chi$. Inserting \eqref{resolution} into the definition of any Feynman integral we find
\be\label{basis-expansion}
\int_{\Gamma} e^{\varepsilon W} \varphi_+ = \la \Gamma \otimes e^{\varepsilon W} | \mathds{1} | \varphi_+ \ra = \la \varphi_i^\vee | \varphi_+ \ra_{\dW} \int_{\Gamma} e^{\varepsilon W} \varphi_i,
\ee
and so the coefficients of a basis expansion on the right-hand side are given by intersection numbers \cite{Mastrolia:2018uzb}. As mentioned previously, these coefficients are always rational functions of kinematic invariants and $\varepsilon$.

Differential equations with respect to external kinematics are actually a special case of the above expansion. To see this, let us expand the action of the differential operator $\partial_{\mu} = \partial / \partial x^\mu$ (for kinematic variables $x^\mu$) on a basis of Feynman integrals:
\be
\partial_{\mu} \int_{\Gamma} e^{\varepsilon W} \varphi_i = \int_{\Gamma} e^{\varepsilon W} (\partial_{\mu} \varphi_i + \varepsilon \partial_{\mu}W \varphi_i).
\ee
The integral on the right-hand side can be treated as the one on the left-hand side of \eqref{basis-expansion} with $\varphi_+ = (\partial_{\mu} + \varepsilon \partial_{\mu}W) \varphi_i$. This gives us straightforwardly
\be
\partial_{\mu} \int_{\Gamma} e^{\varepsilon W} \varphi_i = \la \varphi_j^\vee | (\partial_{\mu} {+} \varepsilon \partial_{\mu}W) \varphi_i \ra_{\dW} \int_{\Gamma} e^{\varepsilon W} \varphi_j. 
\ee
As we did around \eqref{intro-differential}, we can read-off entries of the connection matrix $\mathbf{\Omega}_\mu$ to be
\be\label{Omega}
(\mathbf{\Omega}_\mu)_{ij} =  \la \varphi_j^\vee | (\partial_{\mu} {+} \varepsilon \partial_{\mu}W) \varphi_i \ra_{\dW},
\ee
and so parallel transport on the kinematic space is also governed by intersection numbers \cite{Mastrolia:2018uzb}. If boundary conditions are known, the differential equations can also be used for the practical purpose of expanding Feynman integrals perturbatively in $\varepsilon$, see \cite{Henn:2014qga}.

One can derive higher-order differential equations for a \emph{single} Feynman integral $I = \la \Gamma{\otimes}e^{\varepsilon W} | \varphi_+ \ra$ in a completely analogous fashion. For example, let us take $x$ to be a specific kinematic variable we want to differentiate against. Let us assume that all $\chi$ derivatives $\partial_x^i I$ are linearly independent, i.e., that
\be
\varphi_i = (\partial_x {+} \varepsilon \partial_x W)^i \varphi_+ \qquad i=1,2,\ldots,\chi
\ee
forms a basis of the cohomology group $H_{\dW}^\P$. Then the basis expansion formula gives
\be\label{higher-ode}
I = \sum_{i=1}^{\chi} \la ((\partial_x {+} \varepsilon \partial_x W)^i \varphi_+)^\vee | \varphi_+  \ra_{\dW}\, \partial_x^i I.
\ee
This is an ordinary differential equation of order $\chi$ for the Feynman integral $I$. Based on this differential equation one can infer how complicated the function $I$ is, e.g., distinguish between multiple polylogarithms and elliptic functions. For a recent discussion on this type of bases see \cite{Dlapa:2020cwj}.

In the above manipulations we assumed that one knows how to construct a basis $\la \varphi^\vee_i |$ orthonormal to $| \varphi_j \ra$. In case that such basis is not known up front, it might be easily constructed from an arbitrary non-orthonormal basis $\la \varphi_i |$, which can be always related via a rotation by some matrix $\mathbf{C}^{-1}$,
\be\label{C-inverse}
\la \varphi_i^\vee | = \mathbf{C}_{ij}^{-1} \la \varphi_j |.
\ee
Contracting both sides with $| \varphi_j \ra$ we find that $\mathbf{C}_{ij} = \la \varphi_j | \varphi_i \ra_{\dW}$. Hence the price for not knowing the orthonormal basis is having to compute the inverse of the intersection matrix $\mathbf{C}$ (in other words, $\mathbf{C}$ is in general a non-diagonal metric on the space of twisted forms) \cite{Mastrolia:2018uzb,Frellesvig:2019kgj,Frellesvig:2019uqt}. It would be beneficial to find ways of constructing orthonormal bases from first principles. One approach towards this goal is taken recently in \cite{CHP}. In this work $\la \varphi_i^\vee |$ are chosen to be differential forms supported on cuts, which provides a natural way of making them orthonormal to another set $| \varphi_j \ra$.

Having motivated the need for computing intersection numbers, let us now move on to an overview of methods for evaluating them in practice.

\subsection{\label{sec:evaluating}Evaluating Intersection Numbers}

The complexity of computing \eqref{intersection-number} grows with the number of dimensions of $M$ and the geometry of its boundaries. It will thus be instructive to start with the simplest case when $M$ has one complex dimension ($m{=}1$), which should give the reader an idea of the flavor of techniques involved. Of course, for physical applications we need to understand how to deal with $m{>}1$ cases. In contrast with the one-dimensional one, which is fully understood, computing intersection numbers on general higher-dimensional spaces is an active area of research. We will briefly review one promising strategy in this direction.

\paragraph{One Dimension.} The simplicity here comes from the fact that boundaries can only be points. Since intersection numbers localize on the infinitesimal neighborhood of these points, it is natural to expect that \eqref{intersection-number} will, one way or another, become a sum of residues extracting the local behavior near these boundaries. Let us try to convert this intuition into equations. We start by giving an explicit form of the compactly-supported form $\varphi^c_+$,
\be\label{compact}
\varphi^c_+ = \varphi_+ - \nab\! \sum_{p\in \partial M} \Theta(|z {-} p|^2 {-} \epsilon^2 ) \,\nab^{-1} \varphi_+.
\ee
It is manifestly in the same equivalence class as $\varphi_+$. Here $z$ denotes the complex coordinate on $M$. The sum goes over all points $p$ (orange below) in the boundary $\partial M$ and $\Theta$'s are step functions which are equal to $1$ inside a circle of radius $\epsilon$ around each $p$ (shaded region below) and equal to $0$ outside:
\be
\includegraphics[scale=1,valign=c]{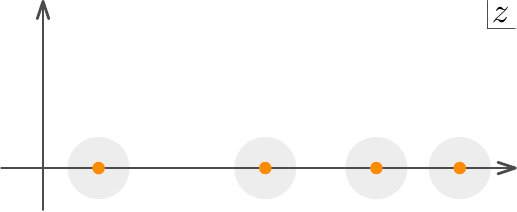}
\ee
One of the reasons for insisting on the use of projective spaces at the beginning is that infinity is now not a special point and can in principle belong to the boundary $\partial M$ on the same footing as other boundary points.
Here $\nab^{-1}$ is a formal inverse of the operator $\nab$ to which we will come back shortly. Note that the step function is non-holomorphic (since a holomorphic function vanishing in some open set also vanishes globally), which is exactly what we need to make the combination $\varphi_- \wedge \varphi_+^c$ receive non-zero contributions.

Acting with $\nab$ on each term in the sum in \eqref{compact} we find
\be\label{varphi-c}
\varphi^c_+ = \varphi_+ \bigg(1- \sum_{p\in \partial M} \Theta(|z {-} p|^2 {-} \epsilon^2 ) \bigg) -  \sum_{p\in \partial M} \delta(|z{-}p|^2 {-} \epsilon^2) \nab^{-1} \varphi_+,
\ee
where we used the fact that $d\Theta$ is a Dirac delta function supported on the circle with radius $\epsilon$ around each $p$. The resulting expression has compact support: the first term vanishes everywhere outside of the small circular neighborhoods of $\partial M$, while the second is supported only on the circles. When inserted into the definition of intersection number,
\be
\la \varphi_- | \varphi_+ \ra_{\dW} = \frac{-\varepsilon}{2\pi i} \int_M \varphi_- \wedge \varphi_+^c,
\ee
the first term in \eqref{varphi-c} does not contribute since $\varphi_- \wedge \varphi_+ = 0$ and we do not have issues with boundaries. Only the second term survives and leaves us with contour integrals
\be
\la \varphi_- | \varphi_+ \ra_{\dW} = \varepsilon \sum_{p \in \partial M} \frac{1}{2\pi i} \oint_{|z{-}p|^2 {=} \epsilon^2} \varphi_- \nab^{-1} \varphi_+.
\ee
This is how localization manifests itself. We recognize each term as a residue, so the final formula reads
\be\label{residue-formula}
\la \varphi_- | \varphi_+ \ra_{\dW} = \varepsilon \sum_{p \in \partial M} \Res_{z=p} \left( \varphi_- \nab^{-1} \varphi_+ \right).
\ee
In order to evaluate this expression we need to find a function $\psi_p = \nab^{-1} \varphi_+$ locally around each point $p$. Acting on both sides with $\nab$, this is just a differential equation for $\psi_p$:
\be\label{de}
\nab\, \psi_p \;=\; \varphi_+ \qquad \text{locally near }z=p.
\ee
The boundary condition is forced upon us by the fact that $\psi_p$ enters a residue formula: it needs to be holomorphic, i.e., $\partial \psi_p / \partial \overbar{z} |_{z=p} = 0$ on $M$. Provided that we keep $\varepsilon$ as a generic non-integer number, a unique holomorphic solution of \eqref{de} exists. It can be found by a holomorphic Laurent expansion of both sides of \eqref{de} and matching the coefficients. Note that we need only a few orders of this expansion which can contribute to the residue in \eqref{residue-formula}. For examples of using this prescription for Feynman integrals on cuts see \cite{Mastrolia:2018uzb,Frellesvig:2019kgj}.

The fact that the solution of $\nab^{-1} \varphi_+$ only exists locally is a rather fundamental issue. As a matter of fact, studying how to ``stitch together'' different local solutions would lead us to a notion of a \emph{sheaf} of such solutions, which gives another way of thinking about cohomology of the loop momentum space. While we will not attempt to explain it here, let us mention that the radius of convergence of \eqref{de} is determined by places where the operator $\nab^{-1} = (d+\varepsilon dW\wedge)^{-1}$ becomes singular. It happens when $dW=0$, i.e., at the critical points we encountered previously in Sec.~\ref{sec:more}. This is not an accident and we will return to it briefly in Sec.~\ref{sec:epsilon}.

\paragraph{Higher Dimensions.} One idea for approaching the computation of intersection numbers on higher-dimensional spaces is to split them into many one-dimensional problems of the type we just encountered above and then ``glue'' the results together. This is actually closely related to what we have been doing all along when we studied gauge theory on the kinematic space. There, prior to loop integration we can think of the loop integrand $e^{\varepsilon W}\varphi_+$ living in the big kinematic space comprising of both external kinematics (schematically $p_i^\mu$) and internal kinematics (schematically $\ell_i^\mu$). We then separate the big space into $p_i^\mu$ and $\ell_i^\mu$ directions and integrate out the latter:
\be
\includegraphics[scale=1,valign=c]{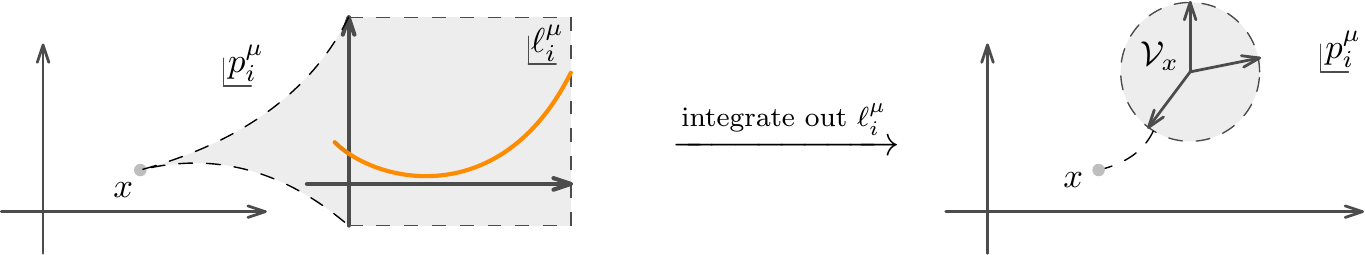}
\ee
The effect of integrating out the internal momenta leaves us with a gauge field (vector bundle) with connection $\mathbf{\Omega}_\mu$ on the external kinematic space. The rank of $\mathbf{\Omega}_\mu$ is computed by the Euler characteristic $\chi$ of the internal momentum space.

The idea is to apply the same strategy by further splitting the internal kinematic space into a ``product'' of many one-dimensional spaces, say labeled by $z_a$ for $a=1,2,\ldots,\P$:
\be
\includegraphics[scale=1,valign=c]{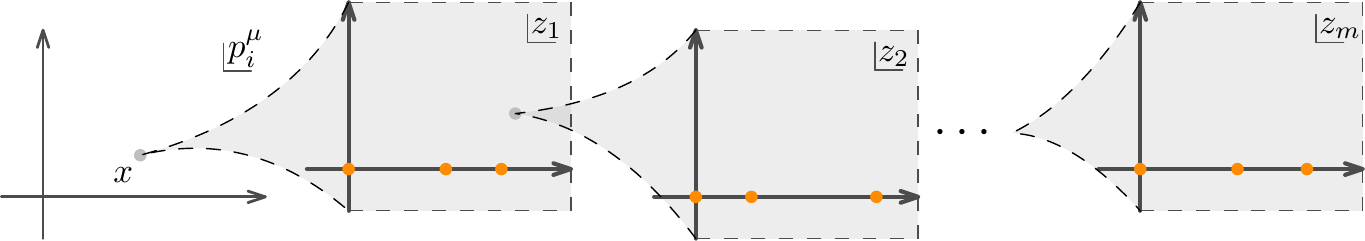}
\ee
Each of these spaces is called a \emph{fiber} and the name for the whole structure is \emph{fiber bundle}, see, e.g., \cite[Ch.~9-10]{Nakahara:2003nw} for introduction. The algorithm for computing intersection numbers goes as follows. At first, the connection on the $\P$-th fiber is given by $\mathbf{\Omega}^{(\P)} = \varepsilon \partial_{z_m}\! W dz_m$, which is just the final component of $\varepsilon dW$. Choosing some basis on the $\P$-th fiber allows one to compute $\mathbf{\Omega}^{(\P-1)}$ on the remaining space using intersection numbers on $z_\P$. It is a $\chi_{\P} {\times} \chi_{\P}$ matrix, where $\chi_{\P}$ is the absolute value of the Euler characteristic of the $\P$-th fiber. Then choosing a basis on the $(\P{-}1)$-th fiber gives a way of computing $\mathbf{\Omega}^{(\P-2)}$ and so on. This gives a recursive way of reaching $\mathbf{\Omega}_\mu$ on the external kinematic space, which is what we were looking for. Since in the intermediate steps we need to compute intersection numbers of involving the higher-rank connections $\mathbf{\Omega}^{(a)}$, one needs to use a ``matrix'' generalization of \eqref{residue-formula}. The final expression involves taking one-dimensional residues and locally solving differential equations in terms of holomorphic expansions. These recursion relations were introduced in \cite{Mizera:2019gea} and applied to Feynman integrals in \cite{Frellesvig:2019uqt}, where we refer the reader for details. Some further improvements are discussed in \cite{CHP,Weinzierl:2020mfq}.

While this algorithm is quite elegant, it also comes with some quirks that need to be understood better. These mostly have to do with the fact that the choice of fibration is highly non-unique and the amount of work in the intermediate steps strongly depends the specific choice of fibers. In particular, there is an implicit assumption on ``genericity'' of each $\mathbf{\Omega}^{(a)}$ (which can be stated in terms of positivity of eigenvalues of $\Res_{z_a=p} \mathbf{\Omega}^{(a)}$ around each boundary point $p$) which is necessary for local solutions of differential equations to exist. A regularization might be necessary if this is not the case. Another possible issue is that for bad choices of fibers the positions of boundaries can be a complicated functions of the coordinates on the remaining fibers. There is clearly a large room for improvements.

\subsection{\label{sec:epsilon}Two Limits}

Complementary to the exact techniques for evaluating intersection numbers from Sec.~\ref{sec:evaluating}, we can attempt to compute them perturbatively in the small dimensional-regularization parameter $\varepsilon$. To gain intuition, let us see how this works in the one-dimensional case first. Since \eqref{de} is a simple ordinary differential equation, its solution near each $z=p$ can be expressed formally as
\begin{align}
\psi_p &= \frac{1}{e^{\varepsilon W(z)}}\int_{p}^{z} e^{\varepsilon W(z')} \,\varphi_+(z') \nn\\
&= \frac{1}{e^{\varepsilon W(z)}} \int_{p}^{z} \left( (z' {-} p)^{\varepsilon \Res_{z'=p}(dW)} + \ldots \right) \left( \frac{\Res_{z'=p}(\varphi_+)}{z'-p} + \ldots \right).\label{psi-p}
\end{align}
In the second equality we isolated the only terms that can contribute to the leading order ${\cal O}(\varepsilon^{-1})$ of the integral. Note that they can only come from the simple pole of $\varphi_+$. Performing the integral in \eqref{psi-p} we find to leading order
\be
\psi_p = \frac{1}{\varepsilon} \frac{\Res_{z'=p}(\varphi_+)}{\Res_{z'=p}(dW)} + \mathcal{O}(\varepsilon^0).
\ee
In particular, at this order $\psi_p$ is a constant in $z$. Therefore the residue in \eqref{residue-formula} is sensitive only to a simple pole of $\varphi_-$, which gives us the final answer to the leading order in $\varepsilon$:
\be\label{epsilon-expansion}
\la \varphi_- | \varphi_+ \ra_{\dW} = \sum_{p \in \partial M} \frac{\Res_{z=p}(\varphi_-)\Res_{z'=p}(\varphi_+)}{\Res_{z'=p}(dW)} + \mathcal{O}(\varepsilon).
\ee
Note that if both $\varphi_-$ and $\varphi_+$ were logarithmic (having only simple poles), then all the steps would have been exact and the result would not have any ${\cal O}(\varepsilon)$ corrections.

There is a second type of perturbative expansion we can make, this time in $\varepsilon^{-1}$. It might seems counterproductive at first, since this means we expand around (negative) infinite space-time dimension, but let us put this interpretation aside for a second. After all, intersection numbers are rational functions, so we can expand them in whichever way is more convenient (with enough perturbative orders one can reconstruct the full result using Pad\'e approximants). One advantage of this approach is that we can expand the differential operator $\nab^{-1}$ in powers of $\varepsilon^{-1}$ (we could not have done it in $\varepsilon$),
\be
\nab^{-1} = \frac{1}{\varepsilon} \frac{1}{dW} + {\cal O}(\varepsilon^0).
\ee
Hence at the order ${\cal O}(\varepsilon^{-1})$ the argument of each residue in \eqref{residue-formula} is the same,
\be
\la \varphi_- | \varphi_+ \ra_{\dW} = \sum_{p \in \partial M} \Res_{z=p} \left( \frac{ \varphi_- \wedge \varphi_+}{dW} \right) + {\cal O}(\varepsilon^{-1}).
\ee
In order not to clutter the notation, we used a convention where the $dz$ forms ``cancel'' between the numerator and denominator, $\frac{dz \w dz}{dz}=dz$. Note that the leading term here is different than the leading term in \eqref{epsilon-expansion}. Moreover, the argument of the residue developed an additional set of poles at $dW=0$, which are the critical points we discussed previously around \eqref{dW}. This is completely expected, because otherwise the answer would have been zero by a residue theorem. We can then deform the contour from enclosing boundary points $p$ (orange) to enclose the critical points $q$ (blue) instead:
\be
\includegraphics[scale=1,valign=c]{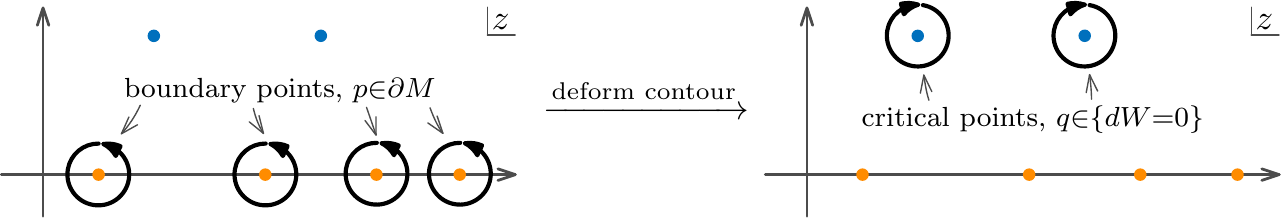}
\ee
The number of critical points is always $\chi$ and that of boundary points is $\chi{+}2$. Using the fact that all zeros of $dW$ are simple, we can evaluate the residues at a cost of a Jacobian:
\be\label{critical-point-formula}
\la \varphi_- | \varphi_+ \ra_{\dW} = -\!\!\!\!\!\!\sum_{q \in \{ dW = 0\}} \!\! \frac{ \widehat{\varphi}_-\,  \widehat{\varphi}_+}{\partial^2 W / \partial z^2} \bigg|_{z=q} + {\cal O}(\varepsilon^{-1}),
\ee
which is yet another localization formula for intersection numbers. Here the hat converts a differential form to a function by stripping away the differential, $\varphi = \widehat{\varphi} \,dz$. Note that since we already established that for logarithmic forms intersection numbers are independent of $\varepsilon$, it must be that ${\cal O}(\varepsilon^{-1})$ cancel out above in those cases.

It is important to understand that this critical-point expansion has little to do with saddle-point approximation for hypergeometric integrals such as \eqref{integral} in the limit $\varepsilon \to \infty$, even though it might look similar. The latter case generically involves a sum over infinitely many critical points on different sheets of the Riemann surface, each of which is weighted by an exponential factor and a phase (for an example, see \cite[App.~A]{Mizera:2019vvs}). By contrast, \eqref{critical-point-formula} is much simpler.

The above patterns persist to higher-dimensional cases. Localization can happen on two distinct sets of points, which we illustrate on the cartoon below:
\be
\includegraphics[scale=1,valign=c]{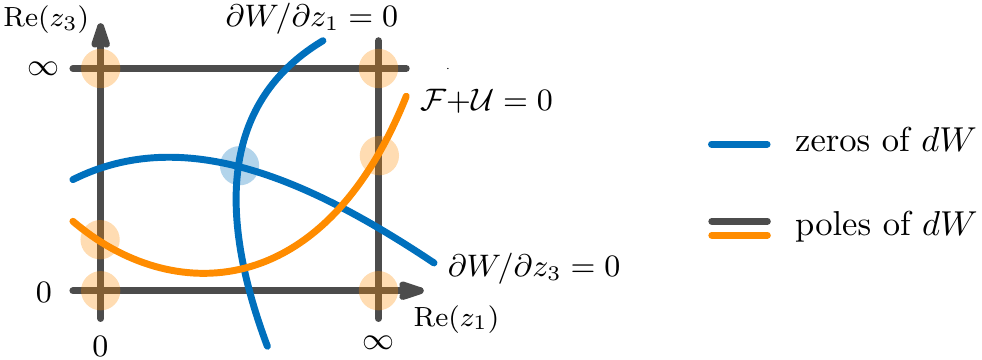}
\ee
The first set (orange) is given by the zero-dimensional components of the boundary $\partial M$. These are the points which lie on the intersection of exactly $\P$ boundary hypersurfaces $H_i$, so we can write $p= H_1 \cap H_2 \cap \cdots \cap H_{\P}$. Here we have to assume that all boundaries intersect normally, that is, there are no points where more than $\P$ hypersurfaces meet. Whenever this is not the case one needs to resolve such a singularity \`a la Hironaka \cite{10.2307/1970486}, in a procedure known as a blow-up (in the context of Feynman integrals, see, e.g., \cite{Bogner:2007cr}). In this sense the above figure, being a two-dimensional slice of an eight-dimensional space, might be misleading, because it cannot possibly show all singularities, such as the non-normally crossing one at $(z_1,z_2,z_3,z_4) = (0,0,0,0)$. After resolving any possible singularities, the generalization of the formula \eqref{epsilon-expansion} is given by
\be\label{boundary-formula}
\la \varphi_- | \varphi_+ \ra_{\dW} = \!\!\!\!\!\sum_{\substack{p \in \partial M \\ p = H_1 \cap \cdots \cap H_{\P}}} \!\!\!\!\! \frac{\Res_p(\varphi_-) \Res_{p}(\varphi_+)}{\prod_{i=1}^{\P} \Res_{H_i} (dW)} + {\cal O}(\varepsilon),
\ee
which is written in terms of higher-dimensional residues, meaning that $\Res_p$ denotes an integral over a product of $\P$ circles surrounding $p$, while $\Res_{H_i}$ is an integral over a tubular neighborhood of $H_i$, see \cite[Ch.~5.1]{griffiths2014principles} for the standard reference. A closed-form formula for higher-order corrections is not currently known, but will most likely involve a sum over not only maximal-codimension boundary components (points $p$), but also next-to-maximal ones at the subleading order, next-to-next in the following order, and so on. The fact that the sum in \eqref{boundary-formula} resembles a Feynman diagram expansion is not an accident and in fact becomes precise when intersection numbers are computed on ${\cal M}_{0,n}$ \cite{Mizera:2019blq}.\footnote{Using similar arguments one can show that the leading $\varepsilon \to 0$ order of a Feynman integral $\la \Gamma {\otimes} e^{\varepsilon W} | \varphi_+ \ra$ is equal, up to an overall constant, to
	\be
	\varepsilon^\P \la \Gamma {\otimes} e^{\varepsilon W} | \varphi_+ \ra = \!\!\!\!\!\sum_{\substack{p \in \partial M \\ p = H_1 \cap \cdots \cap H_{\P}}} \!\!\!\!\! \frac{v_p(\Gamma) \Res_{p}(\varphi_+)}{\prod_{i=1}^{\P} \Res_{H_i} (dW)} + {\cal O}(\varepsilon),
	\ee
	where $v_p(\Gamma)$ equals to $\pm1$ when $\Gamma$ touches the boundary point $p$ (the sign depends on the orientation) and $0$ otherwise.
	Therefore if we choose $\varphi_-$ in a ``dual'' way such that $\Res_p(\varphi_-)= v_p(\Gamma)$ for all $p$, then the formula \eqref{boundary-formula} can be used as a proxy for computing the most divergent part of Feynman integral $\la \Gamma {\otimes} e^{\varepsilon W} | \varphi_+ \ra$. This fact becomes important for coaction properties of Feynman integrals \cite{Abreu:2019wzk}.}

The second set (blue) are simply the critical points determined by $dW=0$. To state the generalization of \eqref{critical-point-formula} we need to assume that all critical points are isolated and non-degenerate. Then the formula reads \cite{Mizera:2017rqa}:
\be\label{critical-higher}
\la \varphi_- | \varphi_+ \ra_{\dW} = (-1)^\P \!\!\!\! \sum_{q \in \{ dW =0\}}  \frac{ \widehat{\varphi}_-\,  \widehat{\varphi}_+}{\det \left( \partial^2 W / \partial z_a \partial z_b\right) }\bigg|_{z=q} +\; {\cal O}(\varepsilon^{-1}).
\ee
The corrections in $\varepsilon^{-1}$ are in principle all known in terms of the so-called \emph{higher residue pairings} \cite{saito1983higher}, see \cite[Sec.~2.4]{Mizera:2019vvs} for detailed expressions. This formula is a double-edged sword: on the one hand, it completely bypasses having to think about the boundary structure of the integration space and any possible blow-ups, but on the other it requires the knowledge of the positions of critical points, which might be difficult or impossible to find analytically. It is rather efficient numerically, which would be particularly useful when combined with finite-field methods.

Note that situations with non-isolated and degenerate points do arise in physical applications at higher-loop orders, see \cite[Sec.~5]{Lee:2013hzt}. It is an open question how to generalize the formula \eqref{critical-higher} to those cases.

Even though both expansions \eqref{boundary-formula} and \eqref{critical-higher} start at the order $\varepsilon^0$, they in general do not agree (just consider $\la \varphi_- | \varphi_+ \ra_{\dW} = \frac{1-\varepsilon}{1+\varepsilon}$). One can ask under what circumstances they do agree. For example, extending the arguments given above one can show that when both $\varphi_-$ and $\varphi_{+}$ are logarithmic then their intersection number is independent of $\varepsilon$ \cite{matsumoto1998} and the two expansions must truncate. This is a sufficient, but not a necessary condition. One reason is that intersection numbers need not be independent of $\varepsilon$ for the two limits to agree, for example $\la \varphi_- | \varphi_+ \ra_{\dW} = \frac{1+\varepsilon + \varepsilon^2}{1+\varepsilon^2}$. Another reason is that twisted forms do not need to be logarithmic for their intersection numbers to be independent of $\varepsilon$, see, e.g., \cite[Sec.~4.1]{Mizera:2019gea} for examples. This is tied to the fact that logarithmicity is not a property of a cohomology class, but rather a twisted form, which is its specific representative.

The above perturbative expansions become particularly useful when applied to the computation of the connection matrix $\mathbf{\Omega}_\mu$, given in \eqref{Omega}. While for a generic choice of the basis $|\varphi_i \ra$, this connection can have a complicated $\varepsilon$-dependence, it was noticed in \cite{Henn:2013pwa} that for a suitable gauge transformation one can often bring it to the simple polynomial form
\be\label{simple-Omega}
\mathbf{\Omega}_\mu =  \mathbf{\Omega}_\mu^{(0)} +  \varepsilon \mathbf{\Omega}_\mu^{(1)},
\ee
where $\mathbf{\Omega}_\mu^{(0)}$ and $\mathbf{\Omega}_\mu^{(1)}$ themselves are $\varepsilon$-independent. Even more so, in the special case when $\mathbf{\Omega}_\mu^{(0)}$ vanishes identically the differential equation for a basis of integrals becomes particularly simple to solve in practice. The corresponding basis is then called \emph{canonical} \cite{Henn:2013pwa}. Note, however, that while \eqref{simple-Omega} is computed with intersection numbers and hence is always a rational function, removing $\mathbf{\Omega}_\mu^{(0)}$ might come at a cost of introducing non-rational functions, see, e.g., \cite{Adams:2018yfj}.

If the connection takes the form \eqref{simple-Omega}, the formulae \eqref{boundary-formula} and \eqref{critical-higher} become especially useful because the expansion truncates and we obtain the full information about the monodromy problem on the kinematic space in just two steps (notice that \eqref{Omega} has terms of order $\varepsilon^0$ and $\varepsilon^1$, so they need to be plugged into the expansions separately). This is particularly striking in the case of the $\varepsilon^{-1}$-expansion, because it means that a finite-length expansion about critical points, which are typically associated to the infinite-dimension limit of Feynman integrals, actually computes the full information about the behavior of the integrals even in four dimensions! For examples, see \cite[Sec.~3.3-3.4]{Mizera:2019vvs}.

Let us stress that a generic basis of Feynman integrals does not lead to the polynomial form \eqref{simple-Omega}. One can ask if there might exist a set of criteria that would allow us to determine whether a given basis leads to such a form ahead of time. For example, we can carry on with the above $\varepsilon$- and $\varepsilon^{-1}$-expansions and once we encounter terms of order ${\cal O}(\varepsilon^2)$ or ${\cal O}(\varepsilon^{-1})$, respectively, we know that the $\varepsilon$-dependence must be non-polynomial. It would be interesting to study such criteria further.

\section{\label{sec:summary}Summary of Open Problems}

In this article we reviewed the status of the connections between Feynman integrals and intersection theory. We use the word \emph{status} in order to emphasize that this research program is still in its early stages and much remains to be understood. To this end, we finish with a non-exhaustive list of open problems (in addition to the ones mentioned throughout the text), which can be freely pursued by interested readers.

\paragraph{Intersection Numbers in the Relative Twisted  Case.} As explained in Sec.~\ref{sec:connection}, the geometry of Feynman integrals in dimensional regularization should be really formulated in terms of relative twisted cohomologies. For example, in the Feynman representation the cohomology would be defined on the manifold $(\CP^1)^\P - \{ {\cal F}{+}{\cal U} =0 \}$ relative to the hyperplanes $\cup_{a=1}^{\P} \{ z_a = 0,\infty \}$ with $\varepsilon$-twisting around the hypersurface $ \{ {\cal F}{+}{\cal U} =0 \}$. As the construction of such cohomology and its dual are rather simple, the question is really about how to compute their intersection pairing, study connections to complex Morse theory, etc. Some progress in this area was made in \cite{matsumoto2018relative} in one dimension. Note that while in this review we talked about the fully twisted case, the older works \cite{FOTIADI1965159,AnalyticStudy1,AnalyticStudy2} mentioned in Sec.~\ref{sec:introduction} deal with the fully untwisted one (in particular, without dimensional regularization). What we are trying to argue is that much would be learned from a synthesis of the two approaches.

\paragraph{Mass Shells and Hypersphere Arrangements.} Other than the Feynman representation, the original loop momentum parametrization mentioned in Sec.~\ref{sec:Feynman} deserves more attention in the context of twisted cohomologies. It has an advantage of having a simple physical interpretation: relative boundaries are mass shells (or lightcones) and the singularity structure is governed by Landau conditions. Moreover, some compact expressions for logarithmic forms have been introduced recently in this representation \cite{ArkaniHamed:2012nw,Arkani-Hamed:2017ahv,Herrmann:2019upk}. It would also be interesting to further explore the connection to hypersphere arrangements for twisted cohomologies \cite{AIF_2003__53_4_977_0,Aomoto:2017npl} in this context.

\paragraph{Analytic Structure of Feynman Integrals.} Following the strategy outlined in Sec.~\ref{sec:introduction}, one can attempt to study the analytic structure of Feynman integrals using intersection numbers more systematically. To be precise, positions of singularities and branch cuts are determined by the positions and degrees of poles of the intersection matrix $\mathbf{\Omega}_\mu$. The latter are most likely highly constrained by the form of the potential $W$ (e.g., given by Symanzik polynomials) and it would be interesting to study the extent to which this dictates the singularity structure on the kinematic space. In a similar spirit one can consider the differential equations \eqref{higher-ode} for individual Feynman integrals, whose singularity structure dictates the class of functions they belong to.

\paragraph{Preferred Bases of Feynman Integrals.} One of the most pressing problems is how to construct orthonormal bases from first principles, i.e., without having to invert a matrix as in \eqref{C-inverse}. Interesting progress on this issue was made recently in \cite{CHP}. In addition, one can ask whether there exists a geometric criterion for bases of twisted forms to produce a canonical system of differential equations. Notice that, unlike individual intersection numbers, for which there are some criteria determining their $\varepsilon$-dependence, the question about canonical differential equations depends on the collective behavior of intersection numbers for the whole basis.

\paragraph{Practical Computations.} For practical applications there is always a need for optimizing efficiency of algorithms computing intersection numbers, which most likely will build upon one of the strategies outlined in Sec.~\ref{sec:intersection}. A particularly interesting direction is combining the resolution of singularities \cite{Bogner:2007cr} with one of those methods. One issue to comment on is that all the algorithms discussed in this article are applicable to general classes of hypergeometric integrals, and as such they do not use the fact that Feynman integrals are not generic integrals, but have an underlying diagrammatic interpretation. This combinatorial nature of Feynman integrals should be exploited one way or another.

\paragraph{Intersection Inception.} In footnote~\ref{sec:aside} we speculated that intersection numbers on the moduli spaces ${\cal M}_{g,n}$ will compute $g$-loop integrands for $n$-point processes in specific quantum field theories (this is not unlikely given that the closely-related ambitwistor string theory has such an interpretation \cite{Geyer:2015bja}). If this was the case, then the resulting intersection number on ${\cal M}_{g,n}$ would define a twisted form on the loop momentum space $M$, and hence the study of the analytic structure on the kinematic space would involve computing intersection numbers of intersection numbers. 

\acknowledgments

This article is based on a talk presented at the workshop ``MathemAmplitudes 2019: Intersection Theory and Feynman Integrals'' held in Padova, Italy on 18-20 December 2019. The author thanks the participants and local organizers of this meeting for creating a stimulating environment. The author gratefully acknowledges the funding provided by Carl P. Feinberg.

%

\bibliographystyle{JHEP}
\bibliography{references}

\end{document}